\documentstyle[12pt,a4,epsf]{article}
\newcommand{\rbibitem}[1]{\bibitem{#1}}
\newcommand{\rcite}[1]{\cite{#1}}
\newcommand{\rref}[1]{\ref{#1}}
\newcommand{\rlabel}[1]{\label{#1}}
\newcommand{\be}{\begin{equation}}
\newcommand{\ee}{\end{equation}}
\begin{document}
\begin{titlepage}
\begin{flushright}
NORDITA - 96/69 N,P\\
hep-ph/9610269
\end{flushright}
\vfill
\begin{center}
{\Large\bf $\pi \to  \l\nu  \gamma$ Form Factors at
Two-loop}\\[1.2cm]
{J. Bijnens and 
P. Talavera \footnote{Research supported by EU under contract number 
ERB 4001GT952585.} }\\[1cm]
{NORDITA, Blegdamsvej 17,\\DK-2100, Copenhagen \O, Denmark}
\end{center}
\vfill
\begin{abstract}
Within Chiral Perturbation Theory (CHPT)
we compute the form factors $A$, $V$ and $\gamma = A/V$ in the
$\pi \to  \l \nu \gamma$ decay to ${\cal O}(p^6)$. $A$ and $\gamma$
obtain corrections of order 25\%.
\end{abstract}
\vfill
October 1996
\end{titlepage}

\section{Introduction}

Interactions of pions at low energies are dominated by the chiral symmetry
of QCD. The best framework for this is Chiral Perturbation Theory as
worked out systematically for the two-flavour case in \rcite{GASSER}. The
assumptions involved here are that the Goldstone Bosons, identified with
the pions, resulting from the
spontaneous breakdown of chiral symmetry are the only relevant degrees
of freedom. We can then expand in the small quark mass and small energies
and momenta. This expansion is good up to a scale of about the chiral symmetry
breaking scale of order of the $\rho$ mass. This expansion works very well in
the purely mesonic sector, some reviews can be found in \rcite{REVIEW}.

In the present article we will deal mainly with the $ SU(2) \otimes SU(2) $
chiral symmetry, where the isospin limit $m_u=m_d$ is taken.
For this case there exist already several full two-loop calculations
\rcite{BGS,pipi2loop,BURGI} and also some calculating only the dispersive parts\rcite{GMeissner,STERN}.
The expansion parameters $\frac{m_\pi^2}{16\pi^2 F_\pi^2}$ and
$\frac{E_\pi^2}{F_\pi^2}$ are quite small but still it is sometimes
useful to go beyond ${\cal O}(p^4)$ because of the following reasons:
\begin{enumerate}
\item The results at ${\cal O}(p^4)$ do not fit the experimental data
       like in the $\gamma \gamma \to \pi^0 \pi^0 $ case\rcite{BGS}.
\item One deal with pions in the final state.In this case
       a isospin zero S-final state produce a strong
       rescattering, and one can obtain huge corrections
       \rcite{pipi2loop,GMeissner}.
\item The quantity under consideration itself only starts at  
${\cal O}(p^4)$
so in order to have an estimate of the accuracy it is necessary to go to 
${\cal O}(p^6)$.
\end{enumerate}
The calculation here is case 1, see section 
\rref{Experiment}, and 3.

There are two problems involved in going beyond ${\cal O}(p^4)$:
\begin{enumerate}
\item The number of constants increases rapidly with the order. 
At ${\cal O}(p^2)$ the lagrangian  contains two constants,
at ${\cal O}(p^4)$  $7$ and at ${\cal O}(p^6)$ there is only a classification for
the three flavour case\rcite{SCHERER}.
\item The algebra involved in such a calculation is rather tedious.
\end{enumerate}

Problem 1 we address by using resonance saturation estimates of the relevant
constants\rcite{TONI}. This is known to work well at ${\cal O}(p^4)$ and was
also used in several other two-loop calculations.

This paper is organized as follows. In Sect. \rref{kinematics}
we define the form factors and discuss the present experimental status.
The next section reviews the presently known theoretical results, which
we have checked independently, about $A$ and $V$. Section 
\rref{FV}
presents our main result, the two-loop calculation of $A$. Technical aspects 
have been placed in two appendices. We then present numerical results
and our conclusions. 
 
\section{Definitions and present experimental results}
\rlabel{kinematics}
A review of the theoretical and experimental situation up to 1982
can be found in \rcite{REPORT}.

\subsection{Matrix element}
We will consider the $\pi^-$ decay
\begin{equation}
\pi^-(p) \to \l^-(p_\l)  \nu(p_\nu) \gamma (q)
\qquad\qquad[\pi_{l2 \gamma}]
\end{equation}
where $ \l $ stands for either $ e $ or $ \mu $ and we will deal with a real
photon, i.e ($q^2=0$).
The process $\pi^+ \to\l^+  \nu \gamma $ can be obtained by
charge conjugation.
The decay with a muon in the final state is completely dominated by
Bremsstrahlung, while this contribution is helicity suppressed in
the electronic mode. This decay therefore presents a good place
to look for the pionic structure.

The decay width can be written as:
\begin{equation}
d\Gamma = \frac{1}{2 M_\pi (2\pi)^5} \sum_{spins} \vert T^2 
\vert d_{LIPS}
(p;p_l,p_\nu,q).
\end{equation}
with
\begin{equation}
d_{LIPS}(p;p_1,\ldots,p_n) = \delta^4(p-\sum_{i=1}^{n} p_i) 
\prod_{i=1}^{n} \frac{d^3p_i}{2p_i^0},
\end{equation}
the phase space volume,
and we have used the covariant normalization of one-particle states:
\begin{equation}
< \vec{p'} \vert \vec{p} > = (2\pi)^3 2 p^0 \delta^3(\vec{p'} - \vec{p}),
\end{equation}
where the matrix element $T$ is calculated inside the $V-A$
theory \rcite{BARDINDAS}:
\begin{equation}
\rlabel{defT}
T = - i G_F e V_{ud}^{*} \epsilon_\mu^{*} \{ F_\pi L^\mu -
H^{\mu \nu} l_\nu \} 
\end{equation}
with:
\begin{eqnarray}
 L^\mu &=& m_l \bar{u} (p_\nu) (1+\gamma_{5}) (\frac{2 p^\mu}{2 p \cdot q}
-\frac{2 p^\mu _l + \not\! q \gamma^\mu}{2 p_l \cdot q}) v(p_l) \nonumber \\
l^\mu &=& \bar{u}(p_{\nu}) \gamma^\mu (1-\gamma_5) v(p_l) \nonumber \\
H^{\mu \nu} &=& i V(W^2) \epsilon^{\mu \nu \alpha \beta} q_\alpha p_
\beta - A(W^2) (q \cdot W g^{\mu \nu} - W^\mu q^\nu) \nonumber\\
W^\mu &=& (p - q )^\mu = (p_l + p_\nu )^\mu.
\end{eqnarray} 
Here $ \epsilon_\mu$ stand for the photon polarization vector with
$q^\mu \epsilon_\mu = 0$, whereas $A$ and $ V$ are the most general
two Lorentz invariant amplitudes occurring in the
decomposition of the tensor
:
\begin{equation}
I^{\mu \nu} = \int \,dx e^{i q \cdot x+i W \cdot y} <0 \vert 
T V^\mu _{em} (x) I^\nu_{4-i 5}(y) \vert \pi^+(p)> , I = V, A.
\end{equation}
at $q^2 = 0$. The most general decomposition can be found in \rcite{DAPHNE}.

As usual we divide the amplitude in two pieces:
\begin{enumerate} 
\item The inner bremsstrahlung  (IB), the photon is emitted by an
external particle.The pion or the lepton in our case. This corresponds to
the term containing $F_\pi L_\mu$ in (\rref{defT}). This is the radiation
of a pointlike electron and pion.
\item The structure dependent part (SD), it is in these terms that the
bound state structure of the pion plays a role. This contains the two
structure functions $V(W^2)$ and $A(W^2)$.
\end{enumerate}
 
Usually the internal Bremsstrahlung
plays the mayor role, masking the structure dependent effects. But in
$\pi\to e\nu\gamma$ it is helicity suppressed, allowing detection
of the structure dependent terms.
A more extended discussion of the kinematics can be found in the review
of particle properties\rcite{PDG} or Ref. \rcite{DAPHNE}.

Time-reversal invariance implies that $A$ and $V$ are real functions for
$W^2$ below the two-pion threshold, which is the region of interest here.
They are analytic functions of $W^2$ with cuts on the positive real axis.

One of the reasons to perform the present calculation is that the $W^2$
dependence of the two form factors only starts at ${\cal O}(p^6)$, see below.

\subsection{Present experimental results}
\rlabel{Experiment}

There are two experiments that have determined $V(W^2)$. They both assumed
a constant form factor. One of the experiments \rcite{EGLI} used
the decay $\pi^+\to e^+\nu e^+ e^-$ thus allowing to determine the
sign of $V$ as well. The average quoted in \rcite{PDG} is
\be
F_V = (0.017\pm 0.008)\,.
\ee
The axial form factor is only measured via the ratio $\gamma = A/V$.
Assuming the CVC value $F_V = 0.0259\pm0.0005$ this yields\rcite{PDG}
\be
F_A = (0.0116\pm0.0016)\,.
\ee
These related to the ones we use by
\be
-\sqrt{2}m_{\pi^+}\left(V,A\right) = \left( F_V, F_A\right)\,.
\ee

There are in the literature some contradictory experimental
results involving the pion decay in a photon and a semileptonic
pair.The last one was reported by Bolotov \rcite{BOLOTOV} .There
they explore a wide kinematical region, observing $80$ $\pi$  decays
in flight, and obtain a result that for the decay width
deviates from the V-A standard model
theoretical calculations by more than three standard deviations.
 This results disagree strongly with previous ones \rcite{BAY}, that
are in agreement with the theoretical calculations (keeping uncertainties).
To obtain a good fit to the data in \rcite{BOLOTOV} 
they introduce
a tensor radiation term of the form:
\begin{equation}
T_{tensorial} = i \frac{e G_F V_{ud}}{\sqrt{2}} \epsilon ^\mu q^\nu
F_T \bar{u}(p_e) \sigma_{\mu \nu} (1+\gamma^5) v(p_\nu) 
\end{equation}
 with a coupling constant $ F_T = - (5.6 \pm 1.7) \times 10^{-3} $. This terms
interferes destructively with the usual $V-A$ form.

The tensor force would be a signal of new physics.
A lot of effort has been devoted to explaining this experiment,
see e.g.
\rcite{VOLOSHIN}, where supersymmetric extensions of
the standard model have been used. This mechanism is
not able to explain the huge disagreement
between the data and the theoretical calculations.
One can try to learn something from similar decays
more accessible experimentally, see e.g. \rcite{CHIZOV}.
Looking at $K^+ \to  \pi^0 \nu e^+ $ as was done in \rcite{AKIMENKO}
could not rule out the possibility of a small tensorial term in the
amplitude.

This motivated us to check whether there were any anomalously large momentum
dependent effects in the form factors. These are not expected \rcite{BARDINDAS}
but a definite calculation of this effect within CHPT was not performed up
to now.

\section{Previous CHPT results}

\subsection{CHPT Lagrangian to ${\cal O}(p^4)$}

We use in this work the sigma model parametrization as used in \rcite{GASSER}.
This uses the equivalence of the groups $SU(2)\otimes SU(2)$ and $O(4)$ and
simplifies the vertices compared to the exponential parametrization.

The most general effective lagrangian consistent with parity,
Lorentz invariance and chiral symmetry is given at lowest
order by:
\begin{equation}
{\cal L}_2 = \frac{F^2}{2} \nabla_\mu U^{\dagger} \nabla_{\mu} U +
2 B F^2 (s^0 U^0 + p^i U^i).
\end{equation}
Where B and F are constants not fixed by symmetry.
We will work here in the Standard CHPT assuming
a large value of $B$.
U(x) denotes a four-component real field in the vector representation of $O(4)$.
s(x) and p(x) are the scalar and pseudoscalar external fields respectively.
We introduce the external sources in the usual way\rcite{GASSER}.
The
covariant derivative is defined as:
\begin{eqnarray}
\nabla_{\mu} U^0 &=& \partial_\mu U^0 + a^i_\mu (x) U^i.\nonumber \\
\nabla_\mu U^i &=& \partial_\mu U^i + \epsilon ^{i k l} v^k_\mu(x) U^l
- a^i_\mu (x) U^0.\nonumber \\
\end{eqnarray}
Here $v^k(x)$ and $a^i(x)$ are the external vector and axial-vector sources.

At ${\cal O}(p^4)$ the Lagrangian is given by
\begin{eqnarray}
{\cal L}_4 &=& l_1 (\nabla^\mu U^\dagger \nabla_\mu U )^2 
+ l_2 (\nabla^\mu U^\dagger \nabla^\nu U)(\nabla_\mu U^\dagger
\nabla_\nu U) \nonumber \\
& &+ l_3 (\chi^\dagger U)^2 + l_4 (\nabla^\mu \chi^\dagger
\nabla_\mu U) + l_5 (U^\dagger F^{\mu \nu } F_{\mu \nu} U) \nonumber \\
& &+ l_6 (\nabla^\mu U^\dagger F_{\mu \nu} \nabla^\nu U) + l_7 (
\tilde{\chi}^\dagger U)^2 + h_1 \chi^\dagger \chi + h_2 F_{\mu \nu}
F^{\mu \nu} \nonumber \\
& & + h_3 \tilde\chi^\dagger \tilde\chi.
\end{eqnarray}
Where the tensor $F_{\mu \nu}$ is defined by:
\begin{equation}
(\nabla_\mu \nabla_\nu - \nabla_\nu \nabla_\mu) U = F_{\mu \nu} U
\end{equation}
which contains the external fields $v_\mu$, $a_\mu$ together with their
derivatives.Also we have introduced the vectors:
\begin{eqnarray}
& & \chi^A = 2 B (s^0,p^i) \nonumber\\
& & \tilde\chi^A = 2 B (p^0, -s^i)
\end{eqnarray}
We have used dimensional regularization in the $\overline{MS}$ scheme.
In this regularization the low-energy constants are defined as:
\begin{eqnarray}
l_i& =& l^r_i + \gamma_i \lambda \qquad( i = 1,\ldots,7) \nonumber\\
\lambda &=& \frac{\mu^{2\omega}}{16 \pi^2} \{ \frac{1}{2\omega}
- \frac{1}{2}(\log(4\pi)+\Gamma'(1) + 1) \}
\end{eqnarray}
where the $l^r_i$ are the coupling constants renormalized at the
scale $\mu$, $\omega = (d-4)/2$
and the $\gamma_i$ factors are found via the Heat-Kernel
expansion and are given in \rcite{GASSER}. The derivative of the $\Gamma$ function is the Euler constant, $\Gamma^\prime(1) = -\gamma$ and
later we will also use $m^2$ for the pion mass squared, $m_\pi^2$.
For the quantities considered here it makes no difference to ${\cal O}(p^6)$ if
we use the full pion mass or only the first term in its quark mass expansion.

\subsection{$A$ form factor at ${\cal O}(p^4)$}

This calculation was performed in Ref. \rcite{GASSER}. The loop contributions,
though allowed by power counting, cancelled and the counterterms were
the only nonzero contribution:
\be
A = \frac{1}{F} \left( 2 l_6 - 4 l_5\right)\,.
\ee
This is in fact the experimental data used to determine this
combination of parameters. It is therefore also useful to have an estimate
of higher order effects to know the accuracy of this determination.

\subsection{$V$ form factor at ${\cal O}(p^6)$}

This part start to play at ${\cal O}(p^4)$ and it involves at least 
one vertex given by the
Wess-Zumino (WZ) lagrangian \rcite{WESS}.
So due to the initial order
at most one-loop diagrams can contribute to ${\cal O}(p^6)$ constructed with
one vertex coming from the WZ Lagrangian and the others from the lowest
order chiral lagrangian (see below). We performed the calculation in the
two-flavour case and our results agree exactly with those
obtained by Ametller et al.\rcite{LLUIS} when restricted to two flavours.
 
The $V(W^2)$ form factor obtained at ${\cal O}(p^6)$ is:
\begin{eqnarray}
V(W^2) &=& -\frac{1}{8\pi^2 F_\pi} \Bigg\{ 1 + \frac{1}{8 \pi^2 F^2}
(- \frac{2}{3} m^2 \log(\frac{m^2}{\mu^2}) \nonumber \\
& &-\frac{4 m^2 - W^2}{6} \bigg(1-\log(\frac{m^2}{\mu^2})-\sqrt{1-\frac{4m^2}{W^2}}
\log(\frac{\sqrt{1-\frac{4m^2}{W^2}}+1}{\sqrt{1-\frac{4m^2}{W^2}}-1})\biggr)
\nonumber \\
& &-\frac{2}{3}m^2+\frac{1}{9}W^2)+\frac{W^2}{m_\rho^2}+... \Bigg\}
\end{eqnarray}
The Kaons are frozen out and the full kaon contribution to  ${\cal O}(p^6)$ 
to this process can
be absorbed in $l_4$, which in turn is absorbed in $F_\pi$.

The divergent parts also agree with those obtained in \rcite{BIJNENS1} for $N_F=2$.
The finite counterterms were estimated by resonance exchange. 
We use the "hidden symmetry" scheme\rcite{BANDO} for the vectors
and calculate the possible
contributions assuming full vector meson dominance.
We have disregarded the contributions of heavier resonances than the $\rho$.
 
Numerical results concerning the $V(W^2)$ form factor can be found in Sect. 
\rref{numerics}.

\section{ $A$ form factor at ${\cal O}(p^6)$ }
\rlabel{FV}

\subsection{Overview}

All quantities have been computed using a Feynman Diagram technique.
We have used two techniques, a brute force one where everything was
calculated and the formfactor $A$ extracted afterwards. This has provided
an independent check of the values for $F_\pi$ and $M_\pi^2$ given in 
\rcite{BURGI}. The other technique was a more aimed extraction of the 
formfactor $A$. This is the one we will describe below.
The $A$ formfactor is the only one that has a contribution proportional to
$g_{\mu\nu} p\cdot q$. 

The presence of the $g_{\mu\nu}$ requires that
the axial-vector insertion and the vector-insertion are in the same one-particle
irreducible subdiagram. This immediately removes a large part of the diagrams.
The presence of the $p\cdot q$ kinematical factor then guarantees it is not
part of the internal Bremsstrahlung contribution.

The direct calculation of the $A$ formfactor has the additional advantage that
some of the most difficult aspects of renormalization at two-loops do
not appear since the ${\cal O}(p^4)$ contribution only was a finite counterterm
contribution. We therefore only need the pion wave function renormalization,
$F_\pi$ and $m_\pi^2$ to ${\cal O}(p^4)$ accuracy. 
This is similar to the situation
in the calculation of $\gamma\gamma\to\pi^0\pi^0$ of \rcite{BGS}.

Another simplifying fact is that in the sigma model parametrization used
here there is only a vector-two-meson vertex in the ${\cal O}(p^2)$ Lagrangian.

A quantity needed in the remainder is
\be
\rlabel{DEFB0}
\frac{m^2}{(d-2)} {\cal B}_0 = \frac{-m^2}{16 \pi^2} \frac{1}{2\omega}
\left\{\left[1+\omega\left(-\log(4\pi)
+\gamma-1\right)\right]_{sub}+
\omega\log\left(\frac{m^2}{\mu^2}\right)\right\}\,.
\ee

\begin{figure}
\begin{center}\leavevmode\epsfxsize=12cm\epsfbox{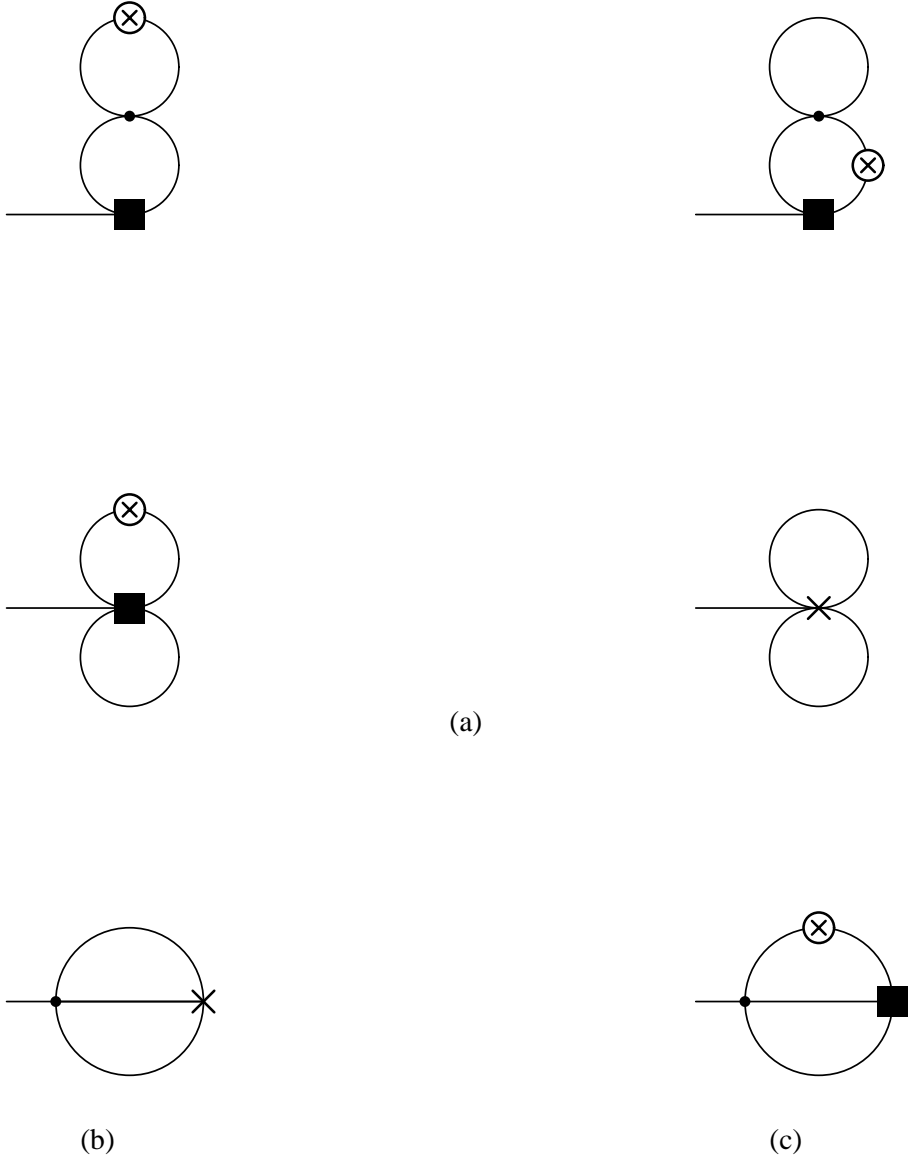}\end{center}
\caption[Two-loop diagrams.]
{\rlabel{figdiag} One particle irreducible two-loop diagrams for $A$.
 The circle-cross
means a vector current, the black box an axial current while
the cross is used for a
vertex with both the  vector and axial current. }
\end{figure}
Let us now discuss the contributions from the remaining diagrams at 
${\cal O}(p^6)$:
\begin{enumerate}
\item Tree level diagrams with one ${\cal O}(p^6)$ vertex. These are needed to
perform renormalization and the finite parts we estimate using resonance
exchange. This is described in subsection \rref{resonance}
\item Tree level diagrams with two ${\cal O}(p^4)$ vertices. 
These can be combined
with the one particle reducible diagrams containing two separate one-loop
subdiagrams and those with a one-loop separated from one ${\cal O}(p^4)$ 
vertex. Together
they combine to applying wave function renormalization to the ${\cal O}(p^4)$ 
result.
They contribute
\be
A = \frac{2 l_6 - 4 l_5}{F} \bigg(1-\frac{1}{F^2} 
\frac{m^2}{(d-2)} {\cal B}_0  \bigg)
\ee
\item The two-loop diagrams with nonoverlapping loops, shown in Fig. 
\rref{figdiag}a
and the one loop diagrams with a ${\cal O}(p^4)$ 
vertex insertion on a propagator
in the loop never produce a factor $p\cdot q$ and hence do not contribute to 
$A$.
\item One-loop diagrams with a ${\cal O}(p^4)$ vertex. 
These are in fact the main 
contribution numerically of the loop diagrams, see section \rref{numerics}.
Analytically they contribute,
\be
A =\frac{1}{F^3} \left(-20l_5+10l_6-8l_1+4l_2\right) \frac{m^2}{(d-2)}
{\cal B}_0
\ee
\item The pure two-loop diagram of Fig. \rref{figdiag}b can never produce 
$g_{\mu\nu}p \cdot q$ and does not contribute to $A$.
\item This leaves now only the diagram in Fig. \rref{figdiag}c.
Its evaluation is
the most difficult part of this calculation.
Its treatment is sketched in App. \rref{appA}.
\item The ${\cal O}(p^4)$ part contains $1/F$, we change this to $1/F_\pi$.
\item The contributions from the counterterms are split into a finite
and an infinite part. The finite part we estimate using resonance
saturation and the infinite part will cancel all the $1/\omega$ divergent terms.
In addition we choose to renormalize such that all
terms $\log(4\pi)-\gamma+1$ disappear as well.
\end{enumerate}

Putting all of these contributions together we obtain:
\begin{eqnarray}
\rlabel{resultA}
\lefteqn{ A(W^2)} &&\nonumber\\&=&
 \frac{1}{F_\pi \vert_{1-loop}} 
 \Biggl\{\left(2l_6-4l_5\right)\bigg(1+\frac{1}{F^2}( l_4 m^2
 + \frac{m^2}{(d-2)} {\cal B}_0
 \bigg)
\nonumber \\
& & + \frac{1}{F^2} \Biggl[ \left(-20l_5+10l_6-8l_1+4l_2\right) 
 ~ \frac{m^2}{(d-2)}{\cal B}_0  
 \nonumber \\
& & -m^2 \Delta(\frac{p \cdot q}{m^2})
\nonumber \\
& & -\frac{m^2}{(16 \pi^2)^2} 
\left(\frac{13}{12}\left(\left[\frac{1}{\omega} -2\log(4\pi)
+2\gamma-2\right]_{sub}+2+2\log\left(\frac{m^2}{\mu^2}\right)\right)
-\frac{307}{5400}\right) \nonumber \\
& &-\frac{p \cdot q}{(16 \pi^2)^2} \left( \frac{1}{18}\left(
\left[\frac{1}{\omega}
-2\log(4\pi)+2\gamma-2\right]_{sub}+2
+2\log\left(\frac{m^2}{\mu^2}\right)\right)+\frac{1427}
{16200} \right) 
\Biggr]  
\nonumber \\ 
& &-\frac{W^2}{m_{a_1}^2} \left(
f_A^2-f_A \alpha_A 2\sqrt{2} \right)
\Biggr\} 
\end{eqnarray}
where ${\cal B}_0$ is defined in (\rref{DEFB0}) and $f_a,\alpha_A$ are 
axial-vector-meson couplings defined in Eq. (\rref{deffalpha}).
We have explicitly shown the single poles, which have to 
be subtracted via counterterms. The renormalization scheme used corresponds to
the one in Ref. \rcite{GASSER}. We subtract in Eqs. (\rref{resultA}) and
(\rref{DEFB0})  the terms in $[\cdots]_{sub}$ and replace $l_4$ by $l_4^r(\mu)$.
The $\Delta(\frac{p \cdot q}{m^2})$ function used here is in Appendix 
\rref{appA}. We have obtained those only
in numerical form. The integrals can be computed rather efficiently.
A full analytical evaluation might be possible similar to $\pi\pi$ scattering
but we have not been able to do all the integrals analytically.

We apply the renormalization
group equation to the chiral lagrangian to obtain one check on our
results \rcite{COLANGELO}.
There are no one-loop contributions proportional to divergent combinations
of the $l_i$. Therefore there are no $1/\omega^2$ poles in the final result.
These cancel as required.

\subsection{Resonance estimates of the ${\cal O}(p^6)$ parameters}
\rlabel{resonance}

We evaluate the resonance contribution following \rcite{TONI}.
Invoking P and C invariance, the relevant lagrangian  can be written
as
\be
{\cal L}_R = \sum_{R=V,A} \bigg\{ {\cal L}_{Kin}(R) + {\cal L}_
{Int}(R) \bigg\},
\ee
with the kinetic terms
\be
{\cal L}_{Kin}(R=V,A)= -\frac{1}{2} \langle \nabla^\lambda R_{\lambda \mu} \nabla_\nu R^{\nu \mu} - \frac{M_R^2}{2} R_{\mu \nu}R^{\mu \nu} \rangle 
\ee
where $\langle {\cal C} \rangle$ means the trace over ${\cal C}$.\\
We have described the vector
and the axial-vector mesons in terms of the antisymmetric tensor fields 
$V_{\mu \nu}$ and  $A_{\mu \nu}$, where we have restricted
ourselves to the octet fields and $M_R$ is the corresponding mass in the chiral limit.\\
The interactions read 
\be
\rlabel{deffalpha}
{\cal L}_{Int}=-\frac{1}{2\sqrt{2}}f_A \langle A_{\mu \nu} F^{\mu \nu}_-
\rangle + i \alpha_A\langle A_\mu [u_\nu,f^{\mu \nu}_+] \rangle 
\ee
All coupling constants are real and we have used
\be
R_{\mu \nu} = d_\mu R_\nu-d_\nu R_\mu
\ee
where the covariant derivative $d_\mu$ acts on the octet multiplets as 
\be
d_\mu R = \partial_\mu R +[\Gamma_\mu,R]
\ee
The connection $\Gamma_\mu$ is defined by
\be
\Gamma_\mu=\frac{1}{2} \{u^\dagger[\partial_\mu-i(v_\mu+a_\mu)]u
+u[\partial_\mu-i(v_\mu-a_\mu)]u^\dagger\}
\ee
The vector-axial field has been defined by:
\be
u_\mu = i u^\dagger D_\mu U u^\dagger = u^\dagger_\mu 
\ee
and the external fields-strength tensor has been introduced via
\be
f^{\mu \nu}_{(\pm)}=u F^{\mu \nu}_L u^\dagger \pm u^\dagger F^{\mu \nu}_R u
\ee
and are associated with the external left and right field sources.

Keeping in mind that the resonance propagator cannot decrease the chiral
counting, the resonance exchange between R, pseudoscalar mesons 
and external fields has to be ${\cal O}(p^6)$.
As input constants we have chosen in the $a_1$ resonance exchange
the following ones \rcite{XIMO}:
\be
\alpha_A  \sim -6.66 ~10^{-3}, \qquad f_A \sim 0.080 \nonumber
\ee
where they have been calculated at leading ${\cal O}(N_C)$.
With these values \rcite{XIMO} quotes a partial width
for the process $\omega \rightarrow \pi^+ \pi^- \pi^0$ of $7.3~MeV$
to be compared with the experimental value $ 7.5 \pm 0.1~MeV$
and a good value for the decay $a_1\to\pi\gamma$.\\
In this case both terms have contribution to the vertex
$\pi^- \rightarrow \gamma a_1$ while only the one coming with
$f_A$ is found in the $a_1 \rightarrow W^-$.\\
There's no $b_1$ contribution due to CP, neither any scalar or tensor
contributes to this order because of spin-isospin.

\section{NUMERICAL RESULTS. CONCLUSIONS}
\rlabel{numerics}

We comment briefly our results about $V(W^2)$ form factor.This is found
to be in perfect agreement with the one found in \rcite{LLUIS}.
We plot in fig. \rref{figV} this form factor versus the lepton
pair invariant mass squared with the photon on mass-shell.
We find that inside the allowed kinematical region the 
variation is around $4.5 \% $.
\begin{figure}
\begin{center}\leavevmode\epsfxsize=12cm\epsfbox{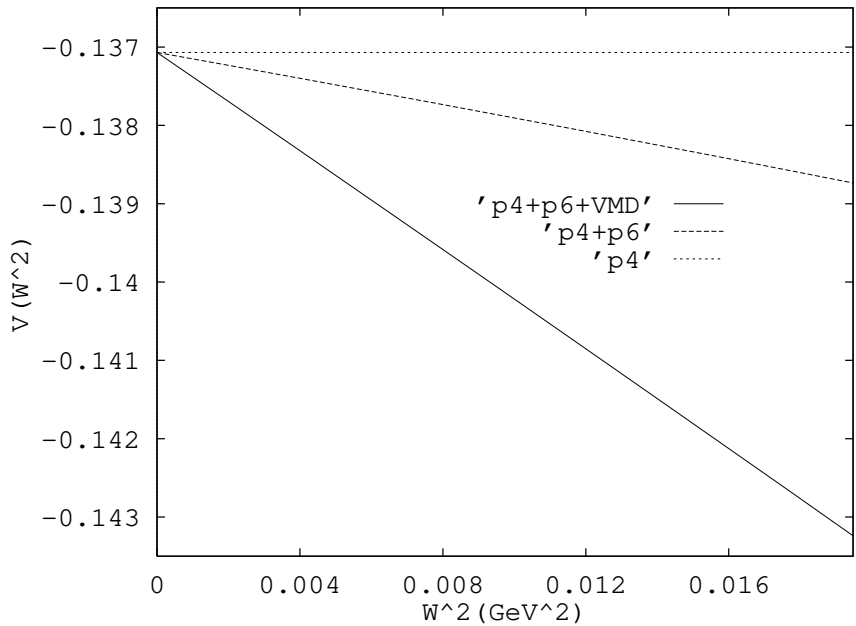}\end{center}
\caption[$V(W^2)$ form factor 
.]
{\rlabel{figV}$V(W^2)$ form factor at ${\cal O}(p^4)$
(short-dashed line),${\cal O}(p^6)$ (dashed line)
and ${\cal O}(p^6)$ plus VMD contribution (full line) vs.
the semi-leptonic pair momenta.}
\end{figure}

For the evaluation of the $A$ form factor we first use the central values
of the $\overline l_r$ quantities in \rcite{GASSER,BCG}.
These are related with the
$l_r$ ones via:
\begin{equation}
l^r_i = \frac{\gamma_i}{32 \pi^2} (\overline l_i + \log(\frac{m^2}
{\mu^2})), ~~i=1,\ldots,6
\end{equation}
At the $m_\rho$ scale these are given by:
\be
\rlabel{standard}
l_1^r = -0.00540\quad l_2^r = 0.00567\quad l_4^r= 0.00560\quad
l_5^r = -0.00553\quad l_6^r = -0.01381\,.
\ee
The other inputs are the charged pion mass and $F_\pi = 0.0924$~GeV.
The size of the various contributions at $W^2 = 0$ can be found
in Table \rref{table1}.
\begin{table}
\begin{center}
\begin{tabular}{|c|c|c|c|}
\hline
$      \mu              $ & $m_\rho$ &0.6~GeV &0.9~GeV  \\
\hline
 ${\cal O}(p^4)$& $- 5.95~ 10^{-2}$ & $- 5.95~ 10^{-2}$ & $- 5.95~ 10^{-2}$\\ 
 $Z_\pi$ and $F\to F_\pi$& $-0.22 ~ 10^{-2}$&
$-0.24  ~ 10^{-2}$&$-0.21 ~ 10^{-2}$ \\ 
 ${\cal O}(p^6)$~ 1-vertex of ${\cal L}_4$& $+1.03~ 10^{-2}$ &
$0.88 ~ 10^{-2}$&$1.19 ~ 10^{-2}$  \\ 
 ${\cal O}(p^6)$ pure two-loops  & $+0.53~10^{-2}$ &
$0.42 ~ 10^{-2}$&$0.59 ~ 10^{-2}$ \\ 
\hline
Total & $-4.62 ~ 10^{-2}$&$ -4.89 ~ 10^{-2}$&$-4.44 ~ 10^{-2}$\\
\hline
\end{tabular}
\end{center}
\caption{Some contributions to the $A(W^2)$ form factor
at $W^2=0$
for $l_i^r(m_\rho)$ of (\protect{\rref{standard}}). Units are $GeV^{-1}$. 
Column 2 is at our standard $\mu$. The other columns show the
variation with $\mu$.
}
\label{table1}
\end{table}
The term proportional to $\Delta$ is very small, about
$5\cdot10^{-6} GeV^{-1}$.
The estimate of the resonance contributions vanishes at $W^2 = 0$. The
pure two-loop diagram provides about 1/3 and the one-loop diagrams with
one vertex of ${\cal L}_4$ about 2/3.
All together the total correction is about 25\%. The remaining subtraction
scale dependence can be judged from comparing the results of the
three scales used in Table \rref{table1}.

The combination of $l_5^r$ and $l_6^r$ that appears here is determined
from $A$ but the variation due to the uncertainty on
$2l_1-l_2$ is rather large.
A better result is obtained by noting that the combination
$2l_1-l_2$ is directly obtainable in the isospin one, spin one $\pi\pi$
channel. It is therefore more suitable to use the experimental
value of $a^1_1$ directly. If we use $a_1^1 = 0.038\pm0.002$ and the value
of $l_4^r$ given above and the ${\cal O}(p^4)$ expression for $a_1^1$ of
\rcite{GASSER} we obtain the range
\be
2 l_1 - l_2 = -0.0172\pm0.0072\,.
\ee
The contribution proportional to $2 l_1-l_2$ from this range
is
\be
A_{2 l_1 - l_2} = (-1.84\pm 0.77)~10^{-2}~GeV^{-1}\,.
\ee
A Roy equation determination of these constants using all available
$\pi\pi$ data\rcite{AB} with $\bar{l}_1 = -1.70\pm0.15 $ and $\bar{l}_2=5.0$
or
\be
\rlabel{ab}
2 l_1 - l_2 = -0.0141\pm0.0003\,.
\ee
This leads to the more restrictive range
\be
A_{2 l_1 - l_2} = (-1.51\pm 0.04)~10^{-2}~GeV^{-1}\,.
\ee
The $W^2$ dependence is rather small and within the range relevant
for this decay extremely linear.
\be
A_{W^2} = 
(-0.0009 -0.00110)\frac{W^2}{m_\pi^2}
\ee
The first term is the two-loop contribution and the second term is the
resonance estimate. The total contribution from the resonance estimate to
the formfactor is rather small, below 2\%.
 
Taking the particle data book values $A=-0.0588\pm0.0081~GeV^{-1}$
and the value in (\rref{ab}) we obtain
\be
\rlabel{liresult}
2 l_5 - l_6 = 0.00315\pm0.00030\,,
\ee
to be compared with the value at ${\cal O}(p^4)$, 0.00275, quoted above.
The quoted error is only the experimental uncertainty. Additional uncertainties
are the $\mu$ dependence (about 5\%), the fact that the other $l_i^r$
are estimated
using $F_\pi = 93.3~MeV$ (a few \%) and contributions from other resonances.
Adding those in quadrature would increase the error in (\rref{liresult})
to about $0.00035$.
We have plotted
the $A$ form factor in Fig. \rref{figA} to show the small $W^2$ dependence.
\begin{figure}
\begin{center}\leavevmode\epsfxsize=12cm\epsfbox{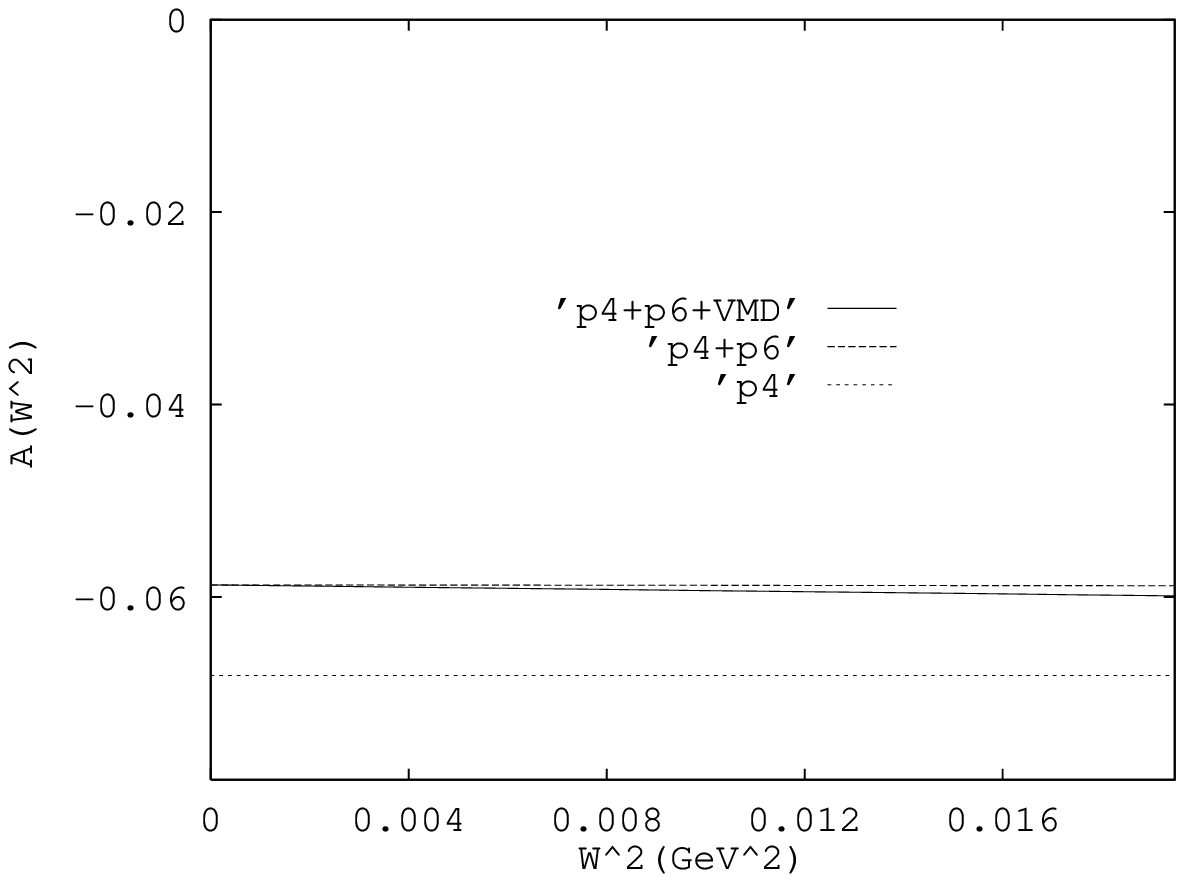}\end{center}
\caption[$A(W^2)$ form factor.]
{\rlabel{figA}The $A(W^2)$ form factor at ${\cal O}(p^4)$ (short-dashed line)
, ${\cal O}(p^6)$ (dashed line) and 
${\cal O}(p^6)$ plus VMD (full line)
vs. the semi-leptonic pair momenta.}
\end{figure}
For completeness we also give the ratio of the two form factors, 
$\gamma = \frac{A}{V}$ in fig. \rref{figG}, together with the present
experimental data.
\begin{figure}
\begin{center}\leavevmode\epsfxsize=12cm\epsfbox{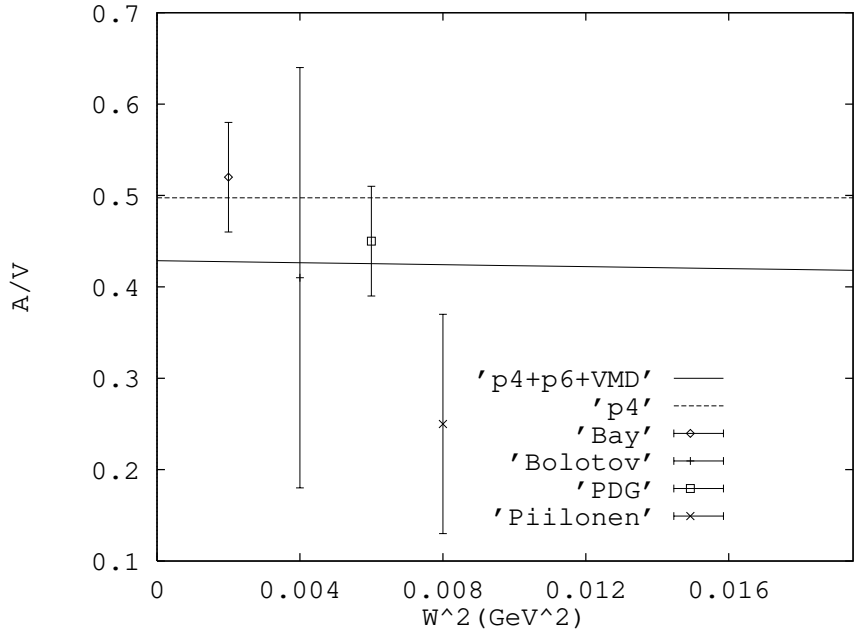}\end{center}
\caption[$\gamma$  factor.]
{\rlabel{figG}The $\gamma = \frac{A}{V}$ factor at ${\cal O}(p^4)$ (dashed line),
 and ${\cal O}(p^6)$+VMD
(full line)
vs. the semi-leptonic pair momenta.We also shown some experimental
results \rcite{PDG},\rcite{BOLOTOV} and \rcite{BAY}}
\end{figure}

The ${\cal O}(p^6)$ corrections to the $A$ form factor are sizable and
diminish it
in absolute value. The $W^2$ dependence is very small as expected and
is dominated by the resonance estimates. The deviation from
the $V-A$ picture observed in \rcite{BOLOTOV} can therefore not
be explained by the small $W^2$ dependence observed here.

The rather large correction found here for $A$ also affects the prediction
of $A$ for the decay $K\to l \nu \gamma$\rcite{DAPHNE}. As a first
guess at the size of the correction we have also calculated the correction
with $M_K$ and $F_K$ used instead of $m_\pi$ and $F_\pi$, 
this results in a 20\% downward correction.
For a definite prediction for the formfactors in the Kaon decay we
have to perform a full $SU(3)$ calculation.

In conclusion, we have performed a two-loop calculation in CHPT for
the $A$ form factor in $\pi\to l\nu\gamma$ and estimated the relevant new constants by resonance
exchange. The corrections are dominated by the loop effects and the
pure two-loop diagram is sizeable. The total correction is about 25\%.
We have also confirmed the known results for the $V$ formfactor.

\section*{Acknowledgements}
We would like to thank the authors of \rcite{INTEGRALS}for providing us with
a preliminary version.
\appendix
\section{Appendix A}
\rlabel{appA}
In this appendix we collect the main functions that enters in the
evaluation of the $A$ form factor at ${\cal O}(p^6)$.

We will take advantage of the fact that we want to extract the part
proportional to $g_{\mu\nu}p\cdot q$. With the notation defined in
App. \rref{appB}, the presence of $g_{\mu\nu}$ requires the presence
of $r_{1\mu}$ and a factor of $r_{1\nu}$ or $r_{2\nu}$ from the other vertex.
We then rewrite $r_1\cdot r_2$, $r_1^2$ and $r_2^2$ in terms of the possible
denominators to put the most divergent parts in terms of integrals with two
denominators only. Those never contribute a factor $p\cdot q$.

The remaining integrals are of five types:
$\langle\langle r_{1\mu}r_{1\nu}\rangle\rangle$,
$\langle\langle r_{1\mu}r_{2\nu}\rangle\rangle$,
$\langle\langle r_{1\mu} r_{1\nu} r_{1\alpha}\rangle\rangle$,
$\langle\langle r_{1\mu} r_{1\nu} r_{2\alpha}\rangle\rangle$ and
$\langle\langle r_{1\mu} r_{2\nu} r_{2\alpha}\rangle\rangle$.
The second and the fourth are trivially related to the first and third respectively by using identities on the $r_2$ subintegral. 
The procedure of evaluation
is described in more detail in App. \rref{appB}.

The function $\Delta(\frac{p \cdot q}{m^2})$ is defined as follows:
\begin{eqnarray}
& &\Delta(\frac{p \cdot q}{m^2}) = \frac{4}{3}
\int_{4m^2}^\infty \frac{[d\sigma]}{\sigma} \int_0^1 \int_0^1 \,dx_1 \,dx_2 x_2
x_1 (1-x_2) 
\bigg\{
x_2(-4+3x_2+4x_2^2) \nonumber \\
& & +\frac{p \cdot q}{m^2} x_2 \bigg(4+6x_2-24x_2^2+x_1(7+10x_2-10x_2^2) 
\bigg) \nonumber \\
& & +\frac{(p \cdot q)^2}{m^4} \bigg(x_2^2(-24+32x_2)+x_1x_2(36-88x_2+44x_2^2)
\nonumber \\ & & +\frac{8}{3} x_1^2x_2^2(2-x_2) \bigg)
\bigg\}
F_3[z]
\end{eqnarray}

As one can see this function make use of the $F_3[z]$ (defined in
Appendix B), which is
finite.
We evaluate this functions numerically.
$\Delta(\frac{p \cdot q}{m^2})$ 
is a  very slowly convergent
function, so one must pay attention in its evaluation.
We have first made a conformal transformation over the $\sigma$ 
variable, which makes the integration over the new
variable simpler. The remaining two integrations ($x_1$, $x_2$) have been evaluated 
in three different ways, and the accuracy of the
results has been compared.
The $\Delta(\frac{p \cdot q}{m^2})$ function can be fit in the region from
0 to 0.5
with a quadratic form with good accuracy,
\be
\left(4\pi\right)^4 \Delta(0\le x \le 0.5) \approx
-0.005648-0.000525 x +0.005195 x^2\,.
\ee

\section{Appendix B}
\rlabel{appB}
This appendix is a minor modification of Refs. \rcite{INTEGRALS,BURGI}
relevant to our case.
For simplicity we have taken in this appendix all the pion masses
normalized to unity.
The multiple space-time integrals will be denoted by
\begin{eqnarray}
& & \langle \ldots \rangle = \int \frac{ \,d^d r_1}{i (2 \pi)^d} ( \ldots), 
\nonumber \\
& & \langle \langle \ldots \rangle \rangle = \int 
\frac{ \,d^d r_1}{i (2 \pi)^d} \int \frac{ \,d^d r_2}{i (2 \pi)^d} ( \ldots). 
\end{eqnarray}
We have also defined the measure as
\begin{equation}
[d\sigma] = \frac{C(w) \Gamma(3/2)}{\Gamma(3/2 + w)} (\frac{\sigma}{4}-1)^
\omega \beta d\sigma, 
\end{equation}
with
\begin{equation}
C(w) = \frac{1}{(4 \pi)^{2+\omega}} ~~,~~ \beta = \sqrt{1-\frac{4}{\sigma}}
\end{equation}
and
\begin{equation}
\lim_{w \rightarrow 0} [d\sigma] = \frac{\beta}{16 \pi^2} d\sigma.
\end{equation}
To show the procedure we will use the vertex diagram in the scalar case:
Given the integral:
\be
V =\langle \langle \prod_{i=1}^4 P_i \rangle \rangle 
\ee
with 
\begin{eqnarray}
& & P_1 = 1-r_1^2 ~~   , ~~  P_2 = 1-(r_1-q)^2 \nonumber \\
& & P_3 = 1-r_2^2 ~~   , ~~  P_4 = 1-(r_1+r_2-p)^2 \nonumber \\
& & p^2 = 1\qquad q^2 = 0.
\end{eqnarray}
We have integrated over one $(r_2)$ of the two 
variables with one-loop techniques.
We subtract the non-local infinities using:
\be
{\cal B}(s) = \overline {\cal B}(s) + {\cal B}(0) 
 = \frac{1}{i} \int \frac{d^dk}{(2 \pi)^d}
\frac{1}{(k^2-1)[(k-p)^2-1]} 
\ee
The finite piece is contained in $\overline {\cal B}$(s) and  we use 
a Cauchy representation for it:
\begin{equation}
\overline {\cal B}(s) = \int_4^\infty \frac{[d\sigma]}{\sigma (\sigma - s)} s
\end{equation}
This yields to:
\begin{equation}
V = \int_4^\infty \frac{[d\sigma]}{\sigma} \langle \frac{s}{P_1 P_2 (\sigma
-s)} \rangle ~,~ s = (p - r_1)^2
\end{equation}
Now we introduce a Feynman parametrization in the way:
\begin{eqnarray}
& &\frac{1}{a_1 \ldots a_n} = 
(n-1)! \int_0^1 \,dx_1 \int_0^{x_1} \,dx_2
\ldots \int_0^{x_{n-2}} \nonumber \\ 
& &\frac{1}{[a_1 x_{n-1}+a_2(x_{n-2}-x_{n-1})
+\ldots + a_n (1-x_1)]^n}
\end{eqnarray}
After this, one is able to write the following expression:
\begin{eqnarray}
& & V = 2 \int_4^\infty \frac{[d\sigma]}{\sigma}  \int_0^1 \,dx_1
\int_0^1 \,dx_2 x_2 \langle \frac{s}{[z-(r_1-R)^2]^3} \rangle, \nonumber \\
& & z = \sigma(1-x_2) + x_2^2 +2 p \cdot q x_2 (1-x_2)(1-x_1) \nonumber \\
& & R = -p(1-x_2)-q x_2(1-x_1).
\end{eqnarray}
With the remaining integrals we have used the following notation:
\begin{eqnarray}
& & \langle \frac{1}{[z-l^2]^m} \rangle = F_m[z] , \nonumber \\
 & & \langle \frac{l_\mu l_\nu}{[z-l^2]^m} \rangle = - \frac{ g^{\mu \nu}}
{2(m-1)} F_{m-1}[z] , \nonumber\\
& & \langle \frac{l_\mu l_\nu l_\alpha l_\beta}{[z-l^2]^m} \rangle = 
\frac{ g^{\mu \nu} g^{\alpha \beta} + cycl.}{4(m-1)(m-2)} F_{m-2}[z].
\end{eqnarray}
Where the function $F_m[z]$ is defined by:
\begin{equation}
F_m[z] = z^{w+2-m} C(w) \frac{\Gamma(m-2-w)}{\Gamma(m)} , m \geq 1
\end{equation}
So the resulting integral can be expressed in terms of:
\begin{equation}
\rlabel{defV}
V_m[P;s] = 2 \int_4^\infty \frac{[d\sigma]}{\sigma} \int_0^1 \,dx_1
\int_0^1 \,dx_2 x_2 P(x_1,x_2) F_m[z]; m = 1,2,3.
\end{equation}
where $P(x_1,x_2)$ is a $x_1$,$x_2$ polynomial.
The $V_m$ integrals can be split in two kinds:
\begin{enumerate}
\item $V_3[P;s]$ is a convergent integral, and can be evaluated using
some gaussian integration subroutine.
\item $V_1[P;s]$ and $V_2[P;s]$ are divergent.We then treat separately.
\end{enumerate}
By partial integration over $x_1$ in (\rref{defV})expression we deal with 
the recursion relation:
\begin{eqnarray}
& & P(x_1,x_2) V_m[z] = P_V(1,x_2) V_m[\overline{z} ] -2 p \cdot q m x_2(1-x_2)
P_V(x_1,x_2) F_{m+1}[z] \nonumber \\
& & P_V(x_1,x_2) = \int_0^{x_1} \,dy P(y,x_2) \nonumber \\
& & \overline z = \sigma(1-x_2)+x_2^2.
\end{eqnarray}
We use it to obtain the maximum number of finite terms.\\
With the rest we have two ways to proceed:
\begin{enumerate}
\item We introduce dimensional regularization and expand 
$\overline z^w$, with this, it is an easy task
to pick up the finite and infinite pieces coming from $V_1[\overline z]$
and $V_2[\overline z]$.
For example in the $V_2[\overline z]$ case we obtain:
\begin{eqnarray}
& & P(x_1,x_2) V_2[z] = 2 C(w) \Gamma(-w) \int_4^\infty
\frac{[d\sigma]}{\sigma} \int_0^1 \,dx_2 x_2 P(1,x_2) \overline z^w 
\end{eqnarray}
working the piece coming from $[d\sigma]$ and keeping the first term
in the $\overline {z}^w$ expansion one can obtain
the result in terms of the $\beta$(n,m) function:
\be
 P(x_1,x_2) V_2[z] = 2 C(w) \frac{\Gamma(-w) \Gamma(3/2)}
{\Gamma(w+3/2)} 4^w \beta(-2w,w+3/2) \nonumber \\
\int_0^1 \,dx_2 x_2 P(1,x_2)
\ee
As one can see those functions contains double poles, that
together with the non-local divergences (with double poles) 
that appear in the subtraction  procedure 
should cancel.
The finite part is evaluated once more in a numerical way.
\item This other possibility is a more aimed one.\\
Consider the divergent integrals containing $V_1[P;s]$ and $V_2[P;s]$.
We have defined the function:
\be
E(m,n) = \int_4^\infty \frac{[d\sigma]}{\sigma} \int_0^1 \,dx (1-x)^m
F_n[\overline{z}], n=1,2.
\ee
By partial integration over the x variable we find the recursion relation:
\begin{eqnarray}
& &(3+\omega+m-n) E(m,n) = \frac{\Gamma(n-2-\omega)}{\Gamma(-\omega) 
\Gamma(n)} Q(\omega+1-n) \\ \nonumber
& &-n \{ E(m,n+1)-E(m+2,n+1)\}
\end{eqnarray} 
where we have defined
\be
Q(\alpha)=C^2(\omega) \Gamma(-\omega) 2 \sqrt{\pi} 4^\alpha
\frac{\Gamma(-1-\omega-\alpha)}{\Gamma(1/2-\alpha)}
\ee
The function E(m,3) is finite at $d=4$, so we use the previous
relation to express $E(m,n) (n=1,2)$ through the divergent
quantities Q and the convergent ones $E(m,3)$.\\
After all one deals with a series in $\omega$ which looks likes:\\
\be
E(i,k) = C^2(\omega) \Gamma^2(-\omega) \{ q(i,k,0) +\omega q(i,k,1)
+\omega^2 q(i,k,2) + \ldots \}
\ee
A full table with the q values can be  found in \rcite{INTEGRALS},
we have checked their values and added some new ones needed for the evaluation
of our integrals.
\end{enumerate}

\vskip 1.5cm
\listoffigures
\end{document}